\documentclass[pra,10pt,twocolumn,showpacs,aps,floatfix,amsmath,amssymb,amsfonts]{revtex4-1}
\usepackage{graphicx}
\usepackage{bm}
\usepackage{subfigure}
\usepackage{color}
\usepackage{amsmath}

\newcommand{\mrm}{\mathrm}

\begin{document}

\title{Quantum defect model of a reactive collision at finite temperature}
\author{Krzysztof Jachymski$^{1}$, Micha{\l} Krych$^{1}$, Paul S. Julienne$^{2}$ and Zbigniew Idziaszek$^{1}$}
\affiliation{$^1$Faculty of Physics, University of Warsaw, Pasteura 5, 02-093 Warsaw, Poland,\\
$^2$Joint Quantum Institute, University of Maryland and National Institute of Standards and Technology, College Park, Maryland 20742, USA}


\date{\today}

\begin{abstract}
We consider a general problem of inelastic collision of particles interacting with power-law potentials. Using quantum defect theory we derive an analytical formula for the energy-dependent complex scattering length, valid for arbitrary collision energy, and use it to analyze the elastic and reactive collision rates. Our theory is applicable for both universal and non-universal collisions. The former corresponds to the unit reaction probability at short range, while in the latter case the reaction probability is smaller than one. In the high-energy limit we present a method that allows to incorporate quantum corrections to the classical reaction rate due to the shape resonances and the quantum tunneling.
\end{abstract}

\maketitle
\section{Introduction}
It has been a long-standing quest in atomic and molecular physics to prepare controllable systems in which inelastic processes at ultralow energies could be studied~\cite{PSJ2012}. Precise control of the internal states and low temperature would give insight into fundamental aspects of quantum physics and chemistry. Much work on this subject has been done using molecular beams~\cite{Rangwala2003,gilijamse2006,Meijer2008,van2012,Narevicius,Jankunas2014,Jankunas2014a}, which finally has lead to recent observation of scattering resonances in Penning ionization using merged beams~\cite{Narevicius}. It also recently became possible to produce ultracold KRb molecules in optical traps by using Feshbach resonance and the STIRAP technique, high phase space density~\cite{Ospelkaus2010,Ni2010,Miranda2011}. A number of different experiments basing on this technique is now being performed with other species, for now mainly consisting of alkali atoms~\cite{Winkler2007,Hudson2008,Deiglmayr2011a,Takekoshi2014}. The electronic, hyperfine, rotational and vibrational state of the produced molecules can be controlled with external fields, so the dependence of chemical reaction rates on the internal state can be analyzed experimentally. Calculations show that many of those molecules can be highly reactive~\cite{Zuchowski2010,Tomza2013}. Another possibility is to study reactions of cold atoms and ions~\cite{Willitsch2008,Rellergert2011,Willitsch2012}. Apart from studying the inelastic collisions, ultracold atoms, ions and molecules offer the opportunity to act as quantum simulators of many-body effects, or to implement quantum information processing protocols~\cite{DeMille2002,Ortner2009,Gorshkov2011,Hazzard2013,Wall2013}.

In the ultracold limit the collision process crucially depends on the quantum statistics, as the scattering of identical fermions exhibits a $p$-wave centrifugal barrier, in contrast to collisions of bosons or distinguishable particles. Quantum effects such as tunneling through the centrifugal barrier play an important role here. Predicting the collisional properties of a complex molecular system is in general a difficult task, in principle requiring precise calculations of potential surfaces~\cite{PSJ2012}. The number of channels and the density of states in molecular collisions can be very high~\cite{Mayle2012,Mayle2013}, making \textit{ab initio} calculations extremely hard. Therefore a need arises for simple theoretical models able to explain experimental results and make predictions on the collision rates.

One class of such models can be built using the formalism of multichannel quantum defect theory (MQDT)~\cite{Seaton1977,Fano,Mies1984a,Mies1984b,Julienne1989,Burke1998,Gao1998b,Gao2001,Gao2005,Gao2008,Idziaszek2009a,Idziaszek2010,Gao2011,Ruzic2013,Gao2013}. This treatment takes advantage of the fact that in many cases the interparticle potential has a known, power-law form ($-C_n/r^n$) at long distances, while the inelastic processes take place only when the particles are very close to each other. The resulting separation of length and energy scales makes the MQDT particularly powerful, allowing to parametrize the short-range physics by the quantum-defect matrix which can be regarded as energy insensitive. If the loss channels are known to have much lesser threshold energies than the entrance channel, the number of parameters needed to describe the scattering process becomes very low. Basing on these ideas, in our previous work~\cite{Jachymski2013} we were able to understand the reaction rates in Penning ionization of Ar by metastable He~\cite{Narevicius} over several orders of magnitude in energy using just two parameters.

In this paper, we provide an extensive description of the results introduced in~\cite{Idziaszek2010,Jachymski2013}. We consider particles which can interact with arbitrary power-law potential $-C_n/r^n$ ($n>3$) at long range. Using a simple model in which the reaction channel has low threshold energy, we derive analytical formulas for the complex scattering length in the entrance channel, from which the elastic and reactive rates can be obtained. They can then be characterized using quantum defect functions, background scattering length and a single parameter which describes the short-range reactivity of the pair of particles. We extend the universal models in which the particles react at short range with unit probability $P^{re}=1$~\cite{Idziaszek2010,Gao2010PRL,Gao2011} to the case when $P^{re}<1$. We analyze in detail the behavior of the collision rates at high and low collision energies. Our results give the correct threshold behavior~\cite{Idziaszek2010R,Quemener2010} as well as classical high-temperature limits~\cite{Langevin1905,Gorin1938}. We discuss the role of tunneling and quantum reflection from the centrifugal barrier and show the corrections to classical results. We then focus on the van der Waals potential ($n=6$), which describes interactions of atoms or molecules without electric or magnetic dipole moments, and describe predictions for the collision rates and the role of shape resonances.

This paper is organized as follows. In Sec. II we briefly review the MQDT formalism. In Sec. III we describe inelastic collisions for isotropic power-law potential at long range using the two-channel quantum defect model. In Sec. IV we consider the case when the exit channel is far below threshold and derive general formulas for the complex scattering length and collision rates using MQDT functions. Sec. V describes the threshold limits for the rates, while Sec. VI focuses on the high energy limits. Finally, in Sec. VII we apply the theory to systems with van der Waals interactions at long range, relating our results to recent experiments. We discuss the results and conclude in Sec. VIII.

\section{Quantum defect formalism}
Our goal is to develop a simple model of a reactive collision which will capture the essential physics. To this end we introduce a multichannel scattering problem, where the internal states are labeled by the index $p$ and the channels by the index $i=\{p\ell m\}$, where $\ell,m$ are the angular momentum quantum numbers, to shorten the notation. We assume that the long-range interaction between the particles is described by a power-law potential $-C_n/r^n$ ($n>3$), with which the characteristic length $R_n = \left(2 \mu C_n/\hbar^2 \right)^{1/(n-2)}$ and energy $E_n = \hbar^2/(2 \mu R_n^2)$ can be associated ($\mu$ is the reduced mass). The short range forces, including the interchannel couplings which are responsible for inelastic processes, are assumed to be limited to the short range, acting at distances $R_0\ll R_n$. The interaction matrix is then asymptotically diagonal
\begin{equation}
W_{pp'}(\mathbf{r})\stackrel{r\to\infty}{\longrightarrow}\left(E^\infty _p+\frac{\hbar^2 \ell(\ell+1)}{2\mu r^2}-\frac{C_n}{r^n}\right)\delta_{pp'}
\end{equation}
where $E^\infty _p$ are the threshold energies for each channel and $\ell$ is the angular momentum quantum number.

We analyze this problem using MQDT, following its formulation by Mies~\cite{Mies1984a,Mies1984b,Mies2000}. In this treatment one first chooses a reference potential $V_p$ in each channel. With each potential one can associate a pair of linearly independent solutions $\hat{f}(r,E)$ and $\hat{g}(r,E)$ that have local WKB-like normalization at short distances
\begin{align}
\label{fghat}
\left.
\begin{array}{lll}
\hat{f}_i(r,E) & \cong & k_i(r)^{-1/2} \sin \beta_i (r),\phantom{\Big(}\\
\hat{g}_i(r,E) & \cong & k_i(r)^{-1/2} \cos \beta_i (r),\phantom{\Big(}\\
\end{array}
\right\}\quad r \gtrsim R_0,
\end{align}
where $k_i(r) = \sqrt{2 \mu \left(E - U_{i}(r)\right)}/\hbar$ is the local wave vector, $U_{i}(r) = V_p(r) + \hbar^2 \ell(\ell+1)/(2 \mu r^2)$, and $\beta_i(r) = \int^r\!\mrm{d}x\,k_i(x)$ is the WKB phase. Another possibility is to use the inhomogeneous Milne equation for parametrization~\cite{Mies2000}. The total wave function at short range can be written as
\begin{equation}
\label{Psi}
\mathbf{\Psi}(r,E)= \mathbf{A}(E) \left[\mathbf{\hat{f}}(r,E) + \mathbf{Y}(E,\ell) \mathbf{\hat{g}}(r,E) \right]
\end{equation}
Here, $\mathbf{A}(E)$ is the amplitude, $\mathbf{\hat{f}}_{ij}=\hat{f}_i\delta_{ij}$, $\mathbf{\hat{g}}_{ij}=\hat{g}_i\delta_{ij}$ are diagonal matrices and $\mathbf{Y}(E,\ell)$ is the so-called quantum defect matrix, a crucial object in this method. At long range the solution of the problem can be expressed using energy normalized functions
\begin{align}
\label{fg}
\left.
\begin{array}{ll}
f_i(r,E) \cong \sin \left(kr-\ell\pi/2+\xi_i\right)/\sqrt{k},\\
g_i(r,E) \cong \cos \left(kr-\ell\pi/2+\xi_i\right)/\sqrt{k},
\end{array}
\right\}\quad r \rightarrow \infty,
\end{align}
where $\xi_i$ denotes the phase shift induced by the full potential $U_i$. The short- and long-range solutions can then be matched using the quantum defect functions $C(E,\ell)$ and $\tan\lambda(E,\ell)$
\begin{align}
\begin{array}{ll}
f_i(r,E) = & C_i^{-1}(E) \hat{f}_i(r)\\
g_i(r,E) = & C_i(E)(\hat{g}_i(r)+\tan \lambda_i(E) \hat{f}_i(r)).
\end{array}
\end{align}
We notice that one can intuitively interpret $C$ and $\tan\lambda$ functions as a measure of deviation of the solution from WKB one. As a result it is clear that for energies high above threshold $C(E)\to 1$ and $\tan\lambda\to 0$.

The solution of any scattering problem is given by the scattering matrix~$S$, from which one can calculate all the relevant quantities~\cite{Taylor1972}. In MQDT framework the $S$~matrix is given in terms of the $R$~matrix~\cite{Mies1984a}
\begin{equation}
\label{Smatr}
\mathbf{S}(E)=e^{i\boldsymbol{\xi}}\left[1+i\mathbf{R}(E)\right]\left[1-i\mathbf{R}(E)\right]^{-1}e^{i\boldsymbol{\xi}},
\end{equation}
where
\begin{equation}
\label{Rmatr}
\mathbf{R}(E)=\mathbf{C}^{-1}(E)\left(\mathbf{Y}^{-1}(E)-\tan\boldsymbol{\lambda}\right)^{-1}\mathbf{C}^{-1}(E)
\end{equation}
Here $\boldsymbol{\xi}_{ij}=\xi_i\delta_{ij}$, $\mathbf{C}_{ij}=C_i\delta_{ij}$ and $\tan\boldsymbol{\lambda}_{ij}=\tan\lambda_i\delta_{ij}$. The pleasing aspect of this theory is that the $\mathbf{Y}(E)$~matrix remains analytic in energy across the thresholds and can usually be regarded as energy- and angular momentum insensitive~\cite{Gao2001}, so that $\mathbf{Y}(E)\approx \mathbf{Y}$ and the matrix elements do not depend on the partial wave. This results from the separation of length and energy scales. As a result, all energy and angular momentum dependence is encoded in $C$ and $\tan\lambda$ functions.

\section{Collision rates in two channel model}
The formalism introduced in the previous section was general and suitable for any multichannel scattering problem. Let us now restrict our attention to a two-channel problem, where $p=1$ is the entrance channel and $p=2$ is the loss channel, both channels are assumed to be open. It is possible to include more channels in the analysis, but this simple case already exhibits interesting features. By choosing the reference potentials to accurately reproduce the scattering lengths in each channel we ensure that $\mathbf{Y}$ contains only off-diagonal terms, $Y_{11}=Y_{22}=0$ and $Y_{12}=Y_{21}=\sqrt{y}$. Under these assumptions we can obtain an analytic formula for the $S$~matrix using formulas~\eqref{Smatr}-\eqref{Rmatr}. The off-diagonal element $S_{1,2}$, which is of particular importance here as it describes the reaction process, is given by
\begin{widetext}
\begin{equation}
\left|S_{1,2}\right|^2=\frac{4yC_1^{-2}C_2^{-2}}{1+2y\left(C_1^{-2}C_2^{-2}-\tan\lambda_1 \tan\lambda_2\right)+y^2\left(C_1^{-2}+\tan\lambda_1\right)\left(C_2^{-2}+\tan\lambda_2\right)}
\end{equation}
\end{widetext}
For $y\ll 1$ this reduces to
\begin{equation}
\left|S_{1,2}\right|^2=4y C_1 ^{-2}(E)C_2 ^{-2}(E).
\end{equation}
The scaling of the loss rate is thus given by the product of $C^{-2}$ functions and the $y$ parameter.

To better understand the meaning of $y$ parameter, one can define a short-range  $S$~matrix as $\mathbf{S}_{\rm sh}=(1-i\mathbf{Y})(1+i\mathbf{Y})^{-1}$ in analogy to Eq.~\eqref{Smatr}, obtaining $S_{11}=\frac{1-y}{1+y}$ and $S_{12}=\frac{2i\sqrt{y}}{1+y}$. This defines the short range reaction probability $P^\mrm{re} = \left|S_{12}\right|^2 = 4 y/(1+y)^2$. $y$ is thus a parameter describing the short-range reactivity and fulfills $0\leq y\leq 1$. The same intuition can be gained by using WKB to find the wave function at short range, as discussed in Section~\ref{apprwkb}.

A convenient way to describe the scattering process is to use energy-dependent complex scattering length, which can be defined as~\cite{Hutson2007,Idziaszek2010}
\begin{align}
\label{alm}
\tilde{a}_{p\ell m}(E) & = \tilde{\alpha}_{p\ell m}(E) - i \tilde{\beta}_{p\ell m}(E) =
\frac{1}{i k} \frac{1-S_{p\ell m,p\ell m}}{1+S_{p\ell m,p\ell m}}.
\end{align}
The elastic and reactive rate constants for channel $p$ are defined as
\begin{align}
\mathcal{K}_p^\mrm{el}(E)=\sum_{\ell,m}{\mathcal{K}^\mrm{el}_{p \ell m}(E)} & = g \frac{h}{2\mu k} \sum_{\ell,m}{\left| 1 - S_{p\ell m,p\ell m}(E) \right|^2}\,,   \label{Keldef} \\
\mathcal{K}_p^\mrm{re}(E)=\sum_{\ell,m}{\mathcal{K}^\mrm{re}_{p \ell m}(E)} & = g \frac{h}{2\mu k} \sum_{\ell,m}{\left(1- |S_{p\ell m,p\ell m}(E)|^2\right)}\,. \label{Klossdef}
\end{align}
Here, $k^2 = 2 \mu E/\hbar^2$ with $E$ denoting the total energy and $g$ is a quantum statistical factor equal to $2$ in the case of identical bosons or fermions in the same internal states, for which only even or odd $\ell$ respectively can occur, or $1$ in other cases. Alternatively, using~\eqref{alm}, we can write
\begin{align}
\mathcal{K}^\mrm{el}_{p \ell m}(E) & = 2 g \frac{h k}{\mu} |\tilde{a}_{p\ell m}(k)|^2 f_{p\ell m}(k) \,,   \label{Kel} \\
\mathcal{K}^\mrm{re}_{p \ell m}(E) & = 2 g \frac{h}{\mu} \tilde{\beta}_{p\ell m}(k) f_{p\ell m}(k) \,, \label{Kloss}
\end{align}
where
\begin{equation}
\label{fellm}
f_{p\ell m}(k) = \frac{1}{1+k^2|\tilde{a}_{p\ell m}(k)|^2+2k \tilde{\beta}_{p\ell m}(k)}.
\end{equation}
Parametrization using $f$ function can be useful, as near threshold we have $f\stackrel{k\to 0}{\rightarrow}1$.

\section{Far from threshold exit channel}
We will now consider the case when the loss channel is strongly open, which means that $E^\infty _2$ is large and negative while we set $E^\infty_1$ to $0$. In this case one can apply the high energy limit for the MQDT functions in the loss channel $C_2(E,\ell)\approx 1$ and $\tan\lambda_2(E,\ell)\approx 0$. The only remaining functions are $C_1$ and $\tan\lambda_1$, so from now on we will drop the index~$1$ in the notation and move the angular momentum dependence to the argument of the functions. Using the analytical results for the $S$ matrix and the definitions from the previous section, we obtain a general formula for the complex scattering length in the entrance channel $\tilde{a}_{\ell m}$
\begin{align}
\label{alm1}
\tilde{a}_{\ell m}(E)  =
- \frac{1}{k}\tan \left[
\xi(E,\ell) - \tan^{-1} \left( \frac{y C^{-2}(E,\ell)}{i + y \tan \lambda(E,\ell)}
\right)\right].
\end{align}
Substituting this into Eqs.~\eqref{Kel} and \eqref{Kloss}, one can express the elastic and reactive rate constants directly in terms of the MQDT functions, obtaining
\begin{equation}
\label{KreMQDT}
\mathcal{K}^\mrm{re}_{\ell m}= g \frac{h}{2 \mu k} P^\mrm{re} \frac{ C^{-2}(E,\ell) (1+y)^2}{(1+y C^{-2}(E,\ell))^2 +y^2 \tan^2 \lambda (E,\ell)}
\end{equation}
for the reactive rate constant and
\begin{widetext}
\begin{equation}
\label{KelMQDT}
\mathcal{K}^\mrm{el}_{\ell m}= g \frac{2 h}{\mu k}\frac{\tan^2 \xi(E,\ell) + y^2 \left(\tan \lambda(E,\ell) \tan \xi(E,\ell) - C^{-2}(E,\ell) \right)^2}{\left(1+\tan^2 \xi(E,\ell)\right) \left( y^2 \tan^2 \lambda(E,\ell) +\left( 1+ y C^{-2}(E,\ell)\right)^2 \right)}
\end{equation}
\end{widetext}
for the elastic rate constant. We note that in contrast to the reactive rate, the elastic one depends explicitly on the phase shift $\xi(E,\ell)$ of the reference potential.

We notice that the properties of the loss channel do not influence the loss rate as long as its threshold energy is far below the threshold energy of the entrance channel, so that the high energy limit can be applied. This observation motivates replacing the problem with the effective single-channel model with a complex potential~\cite{Idziaszek2010}, which gives the same results. Apart from the MQDT functions which depend on the energy, partial wave and the long-range potential, the only remaining parameters in our model are the coupling term $y$ and the phase shift $\xi$ introduced by the full interaction potential. This phase shift determines the background scattering length $a$, which we will express in units of the mean scattering length $\bar{a}$, defined as~\cite{Gribakin1993}
\begin{equation}
\bar{a}=\frac{\pi(n-2)^{(n-4)/(n-2)}}{\Gamma^2\left(\frac{1}{n-2}\right)}R_n.
\end{equation}

\section{Low energy limits}
\subsection{MQDT functions}
We calculated analytically the threshold behavior of MQDT functions directly from their definitions for arbitrary $1/r^n$ potential, extending the previous results derived for $n=6$~\cite{Mies2000} and $n=4$~\cite{NJP2011}. For $s$-wave scattering ($\ell=0$) we obtain
\begin{equation}
C^{-2}(E,\ell=0)\stackrel{E\to 0}{\longrightarrow} k\bar{a}\left(1+(s-\nu)^2\right),
\end{equation}
\begin{equation}
\tan\lambda(E,\ell=0)\stackrel{E\to 0}{\longrightarrow} \nu-s,
\end{equation}
and by definition $\tan\xi\to-k\bar{a}s$, where $s=a/\bar{a}$ and $\nu=\cot\frac{\pi}{n-2}$. Results for the $p$-wave ($\ell=1$), relevant for scattering of ultracold fermions, read
\begin{equation}
C^{-2}(E,\ell=1)\stackrel{E\to 0}{\longrightarrow} k^3 \overline{V} \frac{(1 + (s - \nu)^2) (1 + \nu^2)}{(s - 2 \nu)^2},
\end{equation}
\begin{equation}
\tan\lambda(E,\ell=1)\stackrel{E\to 0}{\longrightarrow} \frac{1+\nu(s-\nu)}{s-2\nu},
\end{equation}
\begin{equation}
\tan\xi(E,\ell=1)\stackrel{E\to 0}{\longrightarrow} k^3 \overline{V} \frac{(1+\nu^2)(1-2s \nu+\nu^2)}{(s-2\nu)(1-3\nu^2)},
\end{equation}
where we have defined the mean $p$-wave scattering volume
\begin{equation}
\overline{V}=\frac{\pi}{9}\frac{(n-2)^{(n-8)/(n-2)}}{\Gamma^2\left(\frac{3}{n-2}\right)}R_n ^3.
\end{equation}
We note that in the case of $n=4$, in the $p$-wave case one has to add the $\propto k^2$ term to the phase shift coming from the long range nature of the potential, as discussed in~\cite{Sadeghpour2000}. This term gives the leading order contribution and modifies the threshold behavior of the elastic rate constant.

\subsection{Elastic and reactive rate}
Formulas derived in the previous section enable us to calculate the low energy limits of the reactive and elastic rate constants using~\eqref{KreMQDT}-\eqref{KelMQDT} and some algebraic transformations. For the reactive rate constant we obtain
\begin{align}
\label{Kl0}
\mathcal{K}^\mrm{re}_{00} & \stackrel{E\to0}{\longrightarrow}  2 g \frac{h}{\mu} \bar{a} y \frac{1+(s-\nu)^2}{1+y^2(s-\nu)^2}, \\
\label{Kl1}
\mathcal{K}^\mrm{re}_{1m} & \stackrel{E\to0}{\longrightarrow}  2 g \frac{h}{\mu } k^2 \overline{V} y \frac{1+\nu^2}{\nu^2} \frac{1+(s-\nu)^2}{y^2(s-\nu+\nu^{-1})^2+(s\nu^{-1}-2)^2},
\end{align}
while for the elastic one
\begin{equation}
\label{Ke0}
\mathcal{K}^\mrm{el}_{00} \stackrel{E\to0}{\longrightarrow}  2 g \frac{h}{\mu} k\bar{a}^2 \frac{s^2+y^2\left(1+\nu^2-s\nu\right)^2}{1+y^2(s-\nu)^2},
\end{equation}
\begin{equation}
\label{Ke1}
\begin{split}
\mathcal{K}^\mrm{el}_{1m} \stackrel{E\to0}{\longrightarrow}  2 g \frac{h}{\mu} k^5 \overline{V}^2 \left(\frac{1+\nu^2}{1-3\nu^2}\right)^2 \times \\ \times \frac{(1-2s\nu+\nu^2)^2+y^2(s+\nu-s\nu^2+\nu^3)^2}{(s-2\nu)^2+y^2(1+(s-\nu)\nu)^2}.
\end{split}
\end{equation}

We note that in the universal regime ($y=1$) all the above formulas reduce to the form which is independent of the $s$ parameter
\begin{equation}
\mathcal{K}^\mrm{re}_{00} \to  2 g \frac{h}{\mu} \bar{a}, \,\,\,\,\, \mathcal{K}^\mrm{re}_{1m} \to  2 g \frac{h}{\mu } k^2 \overline{V}
\end{equation}
\begin{equation}
\mathcal{K}^\mrm{el}_{00} \to  g \frac{h}{2\mu}  (1+\nu^2) k\bar{a}^2, \,\,\,\,\, \mathcal{K}^\mrm{el}_{1m} \to  g \frac{h}{2\mu}\frac{(1+\nu^2)^3}{(1-3\nu^2)^2} k^5 \overline{V}^2.
\end{equation}
Due to the correction coming from the $\propto k^2$ term in the phase shift mentioned above, for $n=4$ the leading term in the $p$-wave elastic rate is proportional to $k^3$ instead.

\section{High energy limits}
\subsection{Reactive rate}
At high energies we first derive an approximate expression corresponding to the classical limit of the scattering. We assume $C^{-2}(E,\ell) =1$ and $\tan \lambda(E,\ell) =0$ for partial waves at which the collision takes place above the barrier, while for collisions below the barrier we take $C^{-2}(E,\ell) =0$. This neglects the effects of the quantum tunneling and of the quantum reflection. In this approximation we obtain
\begin{equation}
\mathcal{K}^\mrm{re} \stackrel{E\to\infty}{\longrightarrow}  \frac{h}{2 \mu} P^\mrm{re} \ell_\mrm{max}(E)\left[1+\ell_\mrm{max}(E)\right]
\end{equation}
where $\ell_\mrm{max}(E)$ is the maximal angular momentum at which the top of the barrier is equal to the collision energy $E$. For a power-law potential $V(r) = - C_n/r^n$ this leads to
\begin{equation}
\mathcal{K}^\mrm{re} \stackrel{E\to\infty}{\longrightarrow}  \frac{h}{2 \mu k} P^\mrm{re} \frac{n}{2} \left( \frac{E/E_n}{\frac{n}{2}-1}\right)^{(n-2)/n}.
\label{KLang}
\end{equation}
In particular we notice that for van der Waals interaction the reactive rate constant behaves as $E^{1/6}$, while for polarization potential ($n=4$) it remains constant at high energy. This can be reproduced by solving the classical problem of scattering on $- C_n/r^n$ potential, assuming that all trajectories that fall on the collision center contribute to the total reaction cross-section $\sigma^\mrm{re}$ with the probability of reaction $P^\mrm{re}$~\cite{Langevin1905}. Then, $\mathcal{K}^\mrm{re} = \sigma^\mrm{re} v$, with $v = \hbar k/\mu$ denoting the mean relative velocity in the gas. We will call this classical result for the reaction rate constant $K^L$. For a general review of high temperature transition state theories, we refer the reader to~\cite{fernandez2006}.

\subsection{Elastic rate}
In the case of the elastic rate constant the situation is not as straightforward, as each partial wave contributes to the elastic cross section. Inspired by the approach of Cote and Dalgarno~\cite{Cote2000}, we derived an approximate expression for the elastic rate constant in the limit of high collision energy. In this approach we consider separately two contributions
\begin{equation}
K^{el} =K^{el,(1)} + K^{el,(2)}.
\end{equation}
where $K^{el,(1)}$, and $K^{el,(2)}$ denote reactive rate from collisions well below the barrier and collisions close or above the centrifugal barrier, respectively. Some characteristic angular momentum $\ell_t$ separating the two regions can be defined such that
\begin{equation}
\label{xitr}
\sin\xi(E,\ell_\mrm{t})=\frac{1}{2}.
\end{equation}
Since this value can be chosen with some flexibility, we decided to pick the value which gives good agreement with numerical calculations. For collisions with angular momenta larger than $\ell_{\mrm{t}}$, we assume that tunneling is not important, thus we neglect the effects of the shape resonances. In this approximation we can set $y=0$ in the formula \eqref{KelMQDT}, obtaining
\begin{equation}
K^{el,(1)} = \frac{2 h}{\mu k} \sum_{\ell > \ell_{\mrm{t}}}(2 \ell +1)
\sin^2 \xi(E,\ell),
\end{equation}
When the collision happens with the energy well below the top of the barrier, one can evaluate the phase shift $\xi(E,\ell)$, using an approximate expression derived in the semiclassical approximation \cite{Bransden1992}
\begin{equation}
\label{semxi}
\xi(E,\ell) \approx - \frac{\mu}{\hbar^2} \int_{r_0}^{\infty} \mrm{d}r \frac{V(r)}{\sqrt{k^2 -(\ell+\frac12)^2/r^2}}
\end{equation}
This formula describes the contribution from the long-range part of the potential $V(r)$, where $r_0$ describes the classical turning point at large distances. In this way for the $1/r^n$ potential we obtain

\begin{equation}
K^{el,(1)}=\frac{\pi^2  h}{8 \mu k} \left(\frac{R_n k}{2\ell_\mrm{t}+1}\right)^{2n-4}\frac{\Gamma(n-1)^2}{(n-2)\Gamma^4(n/2)}.
\end{equation}

In the second regime relevant for collisions close to the top of the barrier or above the barrier we can assume the high-energy limit for MQDT functions, setting $C(E,l)\approx 1$ and $\tan \lambda(E,l)\approx 0$. In principle this approximation works well only for collisions with energies well above the centrifugal barrier. Nevertheless, we make only a small error making a similar approximation for a few partial waves from the region of energies close to the top of the barrier. This yields
\begin{equation}
\label{Kel2}
K^{el,(2)} = \frac{2 h}{\mu k} \sum_{\ell < \ell_{\mrm{t}}}(2 \ell +1) \frac{
\tan^2 \xi(E,\ell) + y^2}{\left(1+\tan^2 \xi(E,\ell)\right) \left( 1+ y\right)^2}.
\end{equation}
In the considered range of angular momenta the phase shifts $\xi(E,\ell)$ vary strongly with the angular momentum. Taking this into account, we can treat $\xi(E,\ell)$ as a random variable in this regime, and we can perform an average assuming uniformly distributed phase shifts
\begin{equation}
\frac{1}{\pi} \int_0^\pi d \xi \frac{
\tan^2 \xi + y^2}{\left(1+\tan^2 \xi\right) \left( 1+ y\right)^2} = \frac{1+y^2}{2(1+y)^2}
\end{equation}
Substituting this into \eqref{Kel2} we get
\begin{equation}
K^{el,(2)}=\frac{2h}{\mu k} \frac{1+y^2}{2(1+y)^2} \left(\ell_\mrm{t} +\frac12\right)^{2}.
\end{equation}

In order to calculate $\ell_\mrm{t}$ we can use the semiclassical expression \eqref{semxi} again, substituting it into \eqref{xitr}, which gives
\begin{equation}
2\ell_\mrm{t}+1=\left( \frac{3\Gamma(n-1)}{\Gamma^2(n/2)} \right)^{1/(n-1)}\left(\frac{E}{E_n}\right)^{(n-2)/(2n-2)}.
\end{equation}
This finally yields the result for the total elastic rate constant
\begin{equation}
\begin{split}
\label{Kelhigh}
K^{el} \approx &\frac{h R_n}{\mu} \left( \frac{\pi^2}{32(n-2)} +\frac{1+y^2}{4(1+y)^2} \right)\\
&\times\left( \frac{3 \Gamma(n-1)}{\Gamma^2(n/2)} \right)^{\frac{2}{n-1}}\left( \frac{E}{E_n} \right)^{\frac{n-3}{2n-2}}.
\end{split}
\end{equation}

\subsection{Approximate treatment including shape resonances}\label{apprwkb}

\begin{figure}
\includegraphics[width=\linewidth]{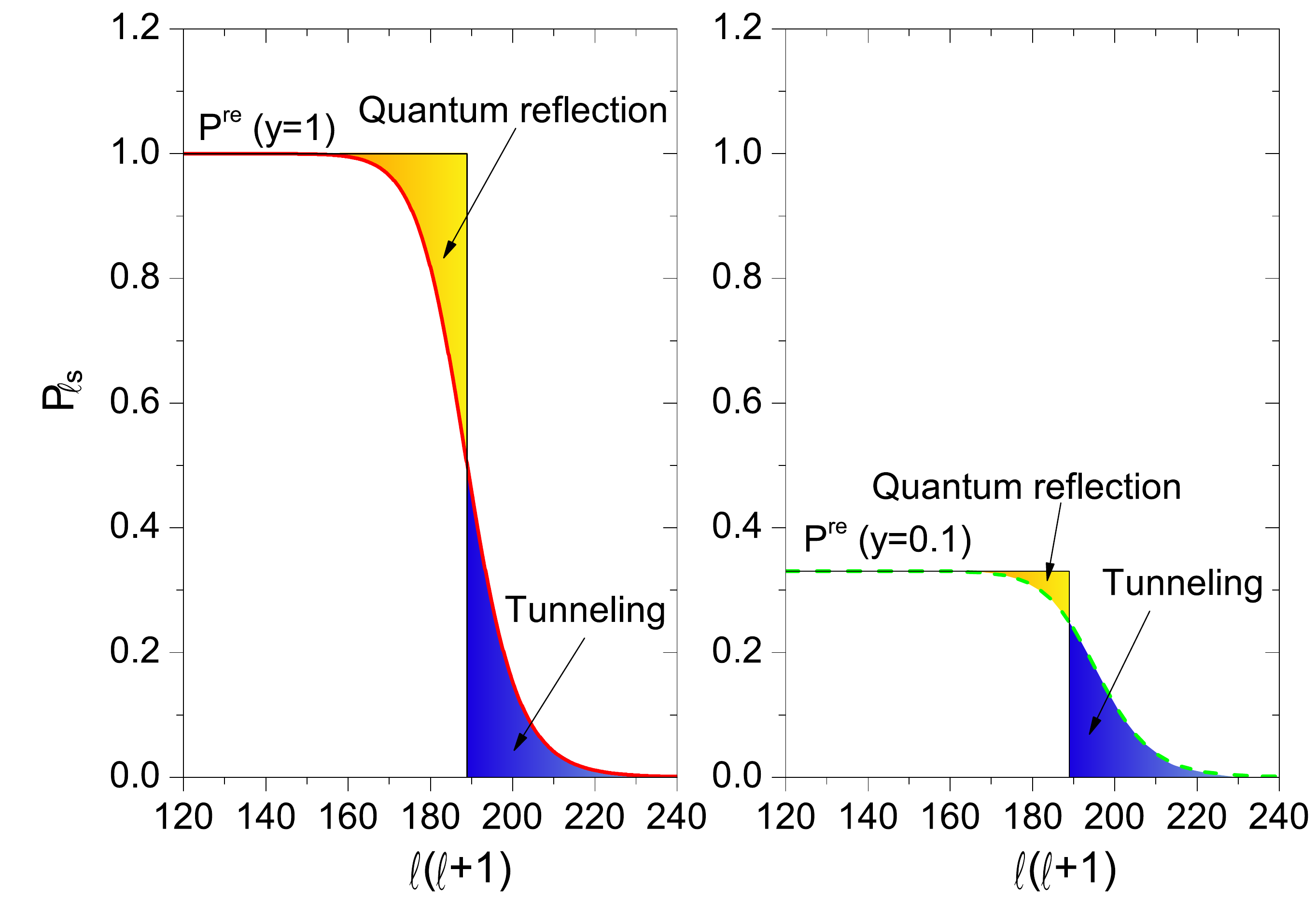}
\caption{
\label{Ferm}
(color online) Reaction probability calculated for a parabolic potential fitted to the actual centrifugal barrier of the van der Waals potential versus the angular momentum squared $\ell(\ell+1)$ (red solid dashed and green dashed lines). The result is independent of the short-range parameter $s$, and is averaged over the short-range phase. Langevin approximation depicted by black solid lines, assumes constant reaction probability $P^\mrm{re}$ above the barrier, and no reaction below the barrier.}
\end{figure}

\begin{figure}
\includegraphics[width=0.9\linewidth]{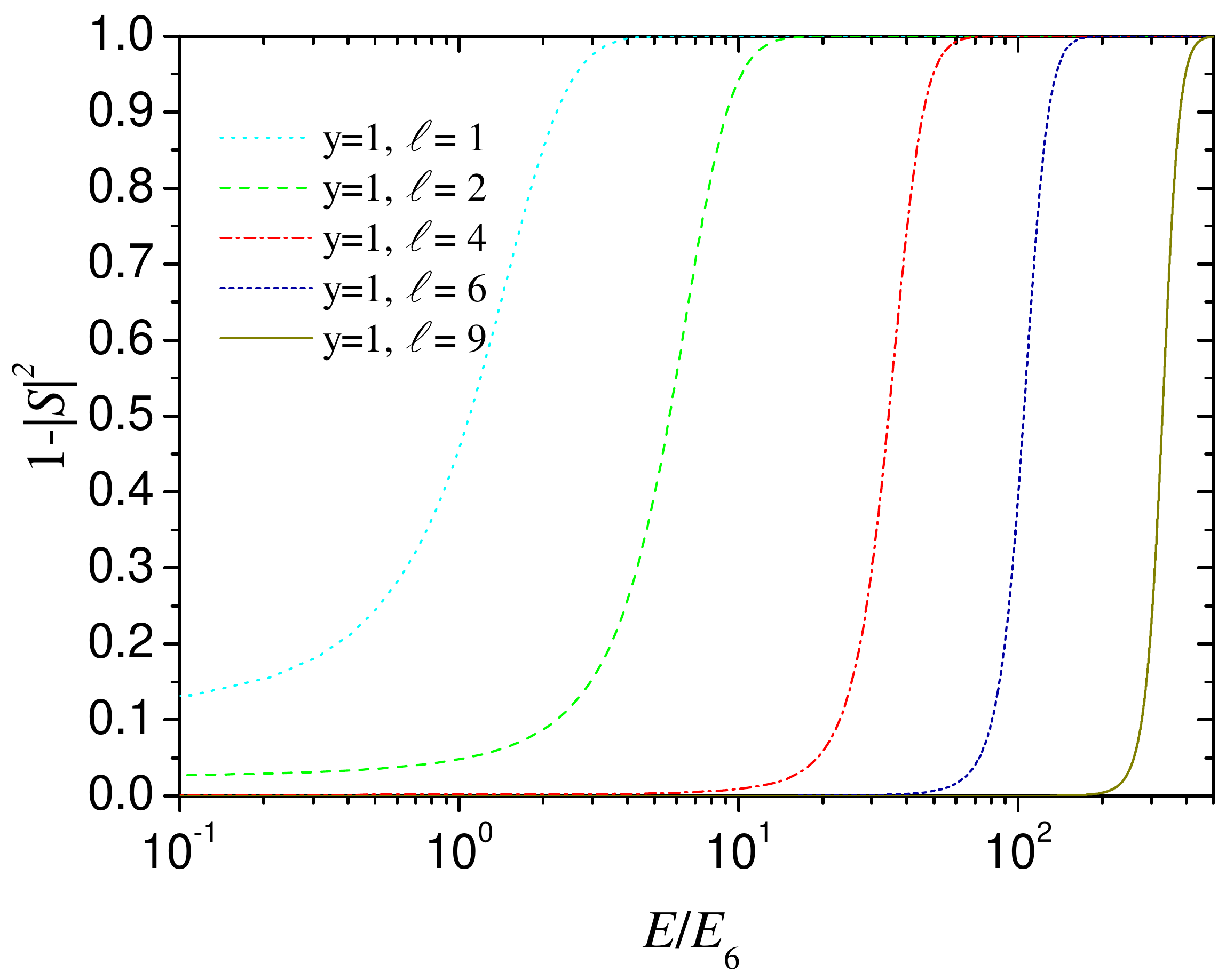}
\includegraphics[width=0.9\linewidth]{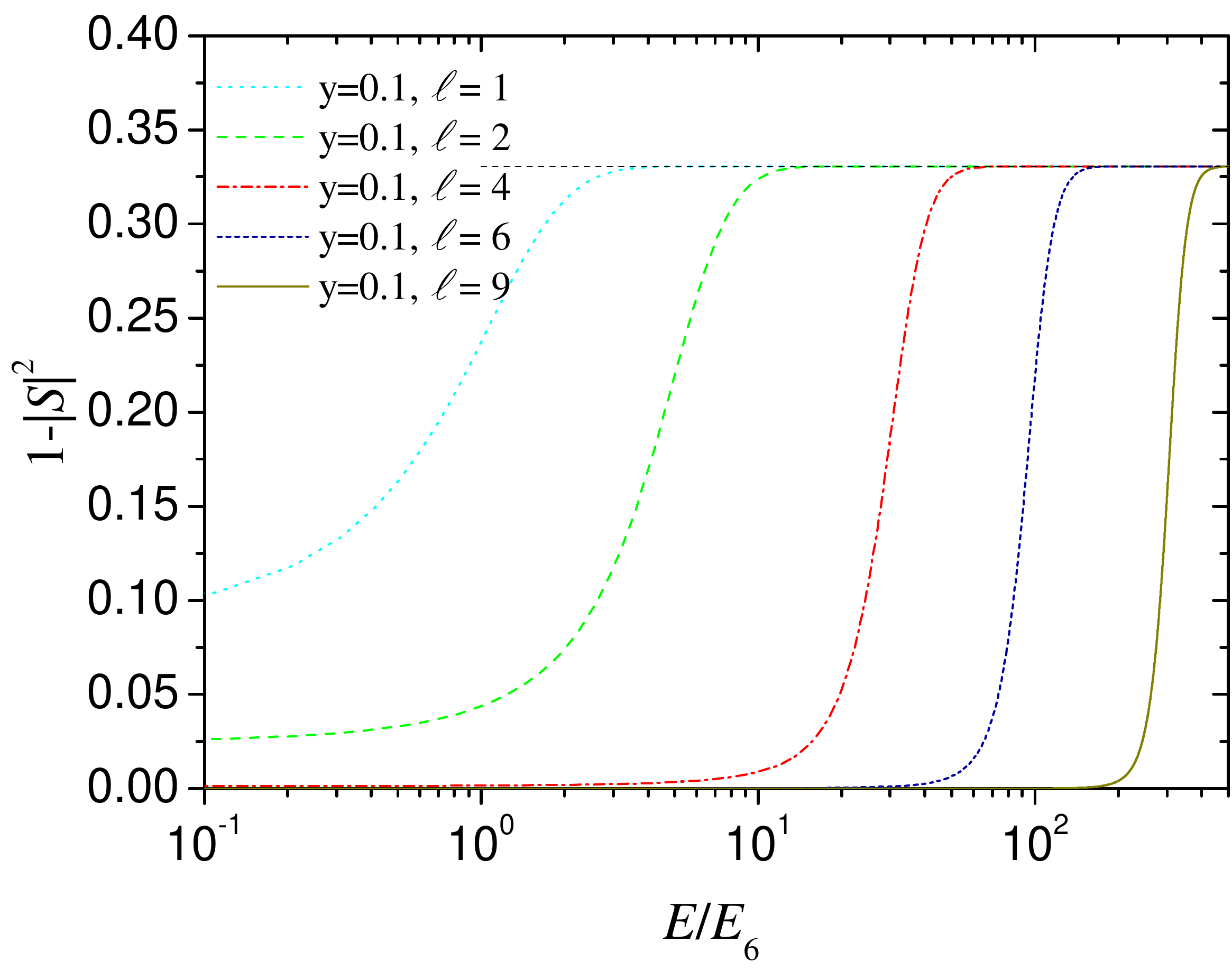}
\caption{\label{partial}(color online) Contribution of few partial waves to reaction rates for universal ($y=1$, top) and nonuniversal case ($y=0.1$, bottom) within the parabolic approximation for the centrifugal barrier averaged over the short range phase. Horizontal black line represents $P^\mrm{re}$}.
\end{figure}
\begin{figure}
\centering
\subfigure[van der Waals potential]{
   \includegraphics[width=0.5\textwidth]{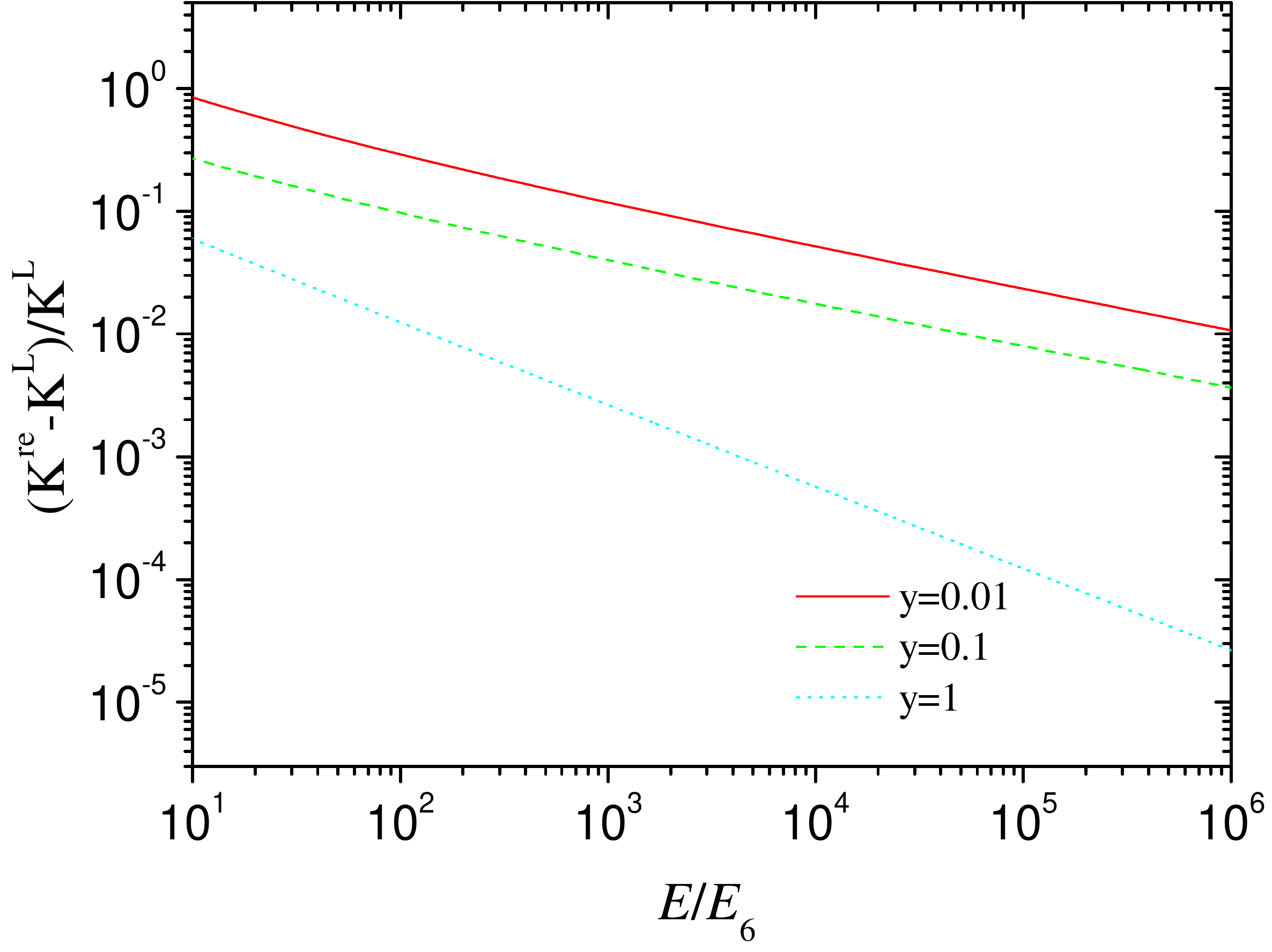}
   \label{fig:subfig1}
 }
\subfigure[Polarization potential]{
   \includegraphics[width=0.5\textwidth]{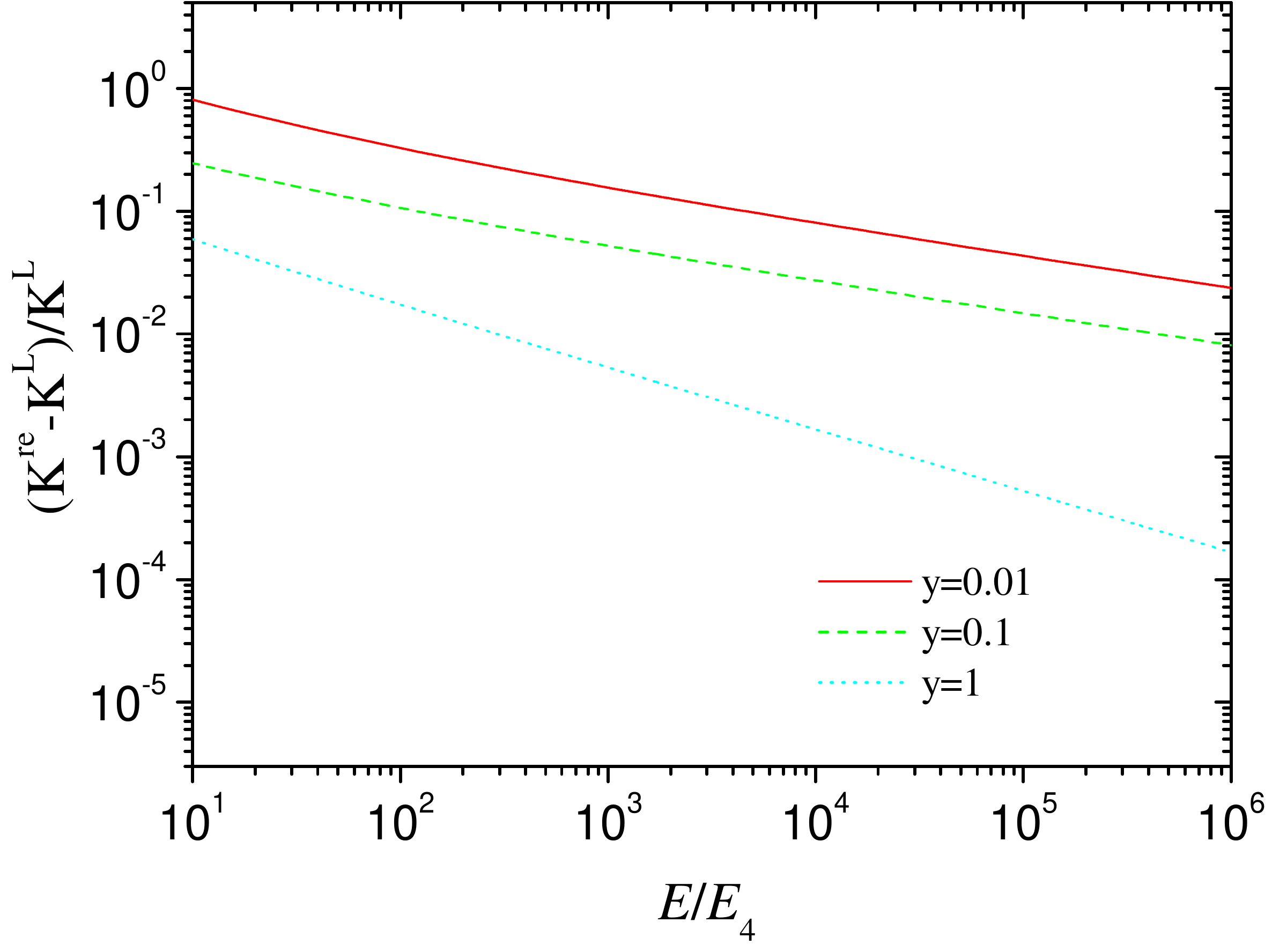}
   \label{fig:subfig2}
 }
\caption{
\label{myfigure}
(color online) Quantum corrections to the classical reaction rates due to the contribution of the shape resonances. Presented results are averaged over the short-range phase, as its value becomes unimportant at large energies.}
\end{figure}

\begin{figure}
\centering
\subfigure[\quad $y=0.1$]{
   \includegraphics[width=\linewidth]{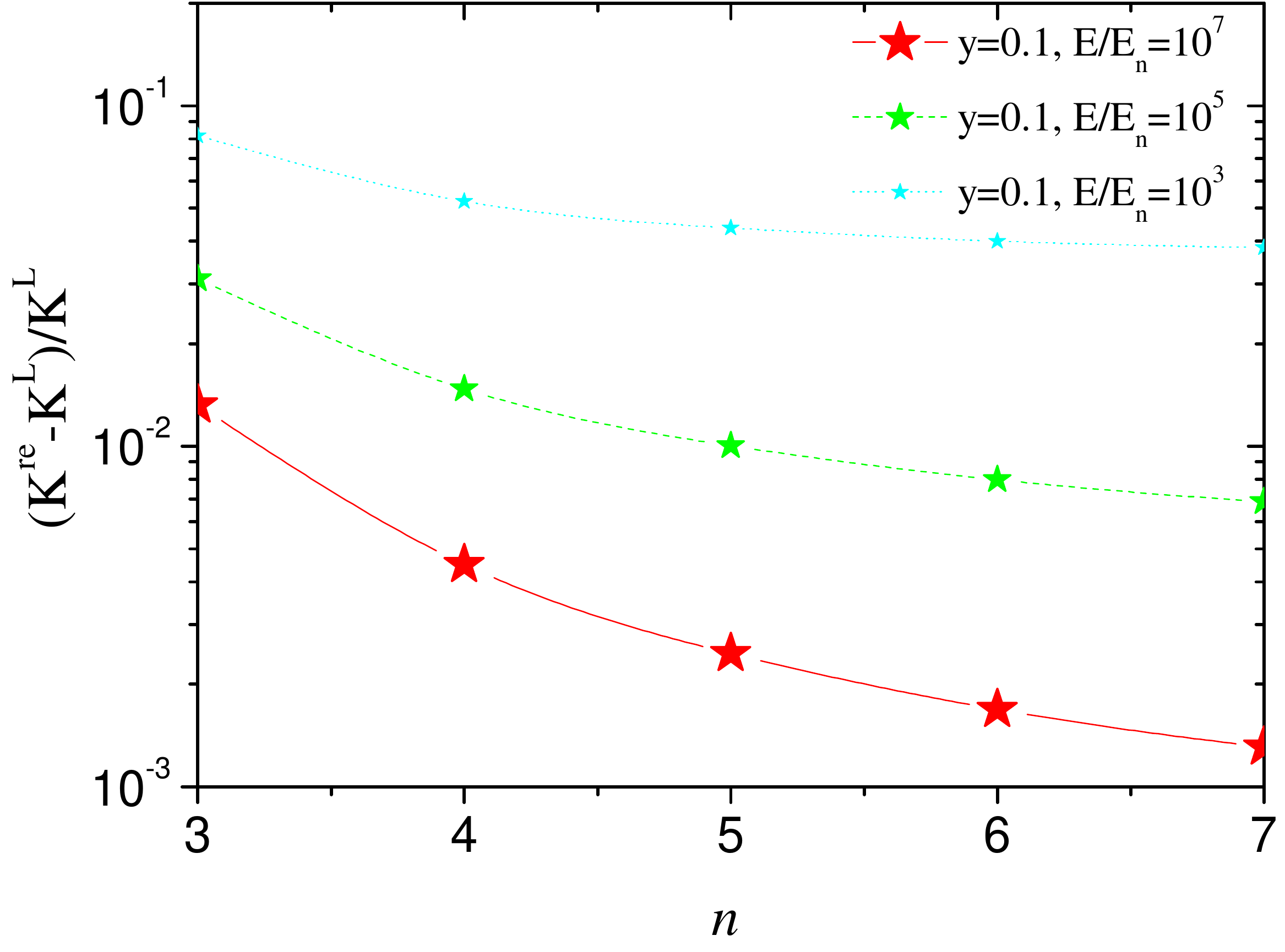}
   \label{fig:subfig5}}

\vspace{0.5cm}
\subfigure[\quad $y=0.01$]{
   \includegraphics[width=\linewidth]{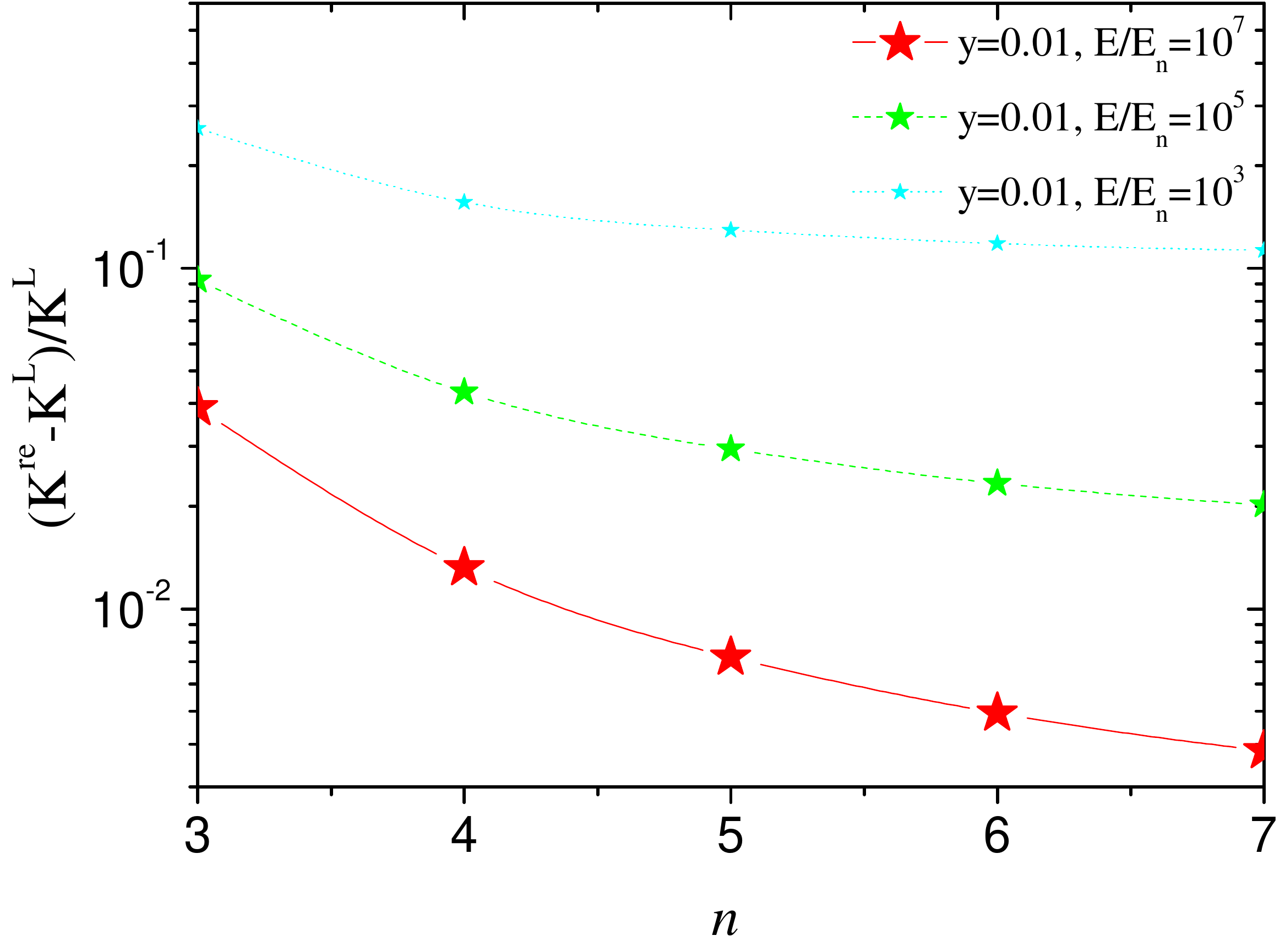}
   \label{fig:subfig6}}
\caption{
\label{myfigure2}
Relative corrections of reaction rates given by the quantum analytical model with respect to the classical approach, calculated for different interaction potentials $1/r^n$, energies and amplitudes of reactions $y$.}
\end{figure}

\begin{figure}
\centering
\includegraphics[width=\linewidth]{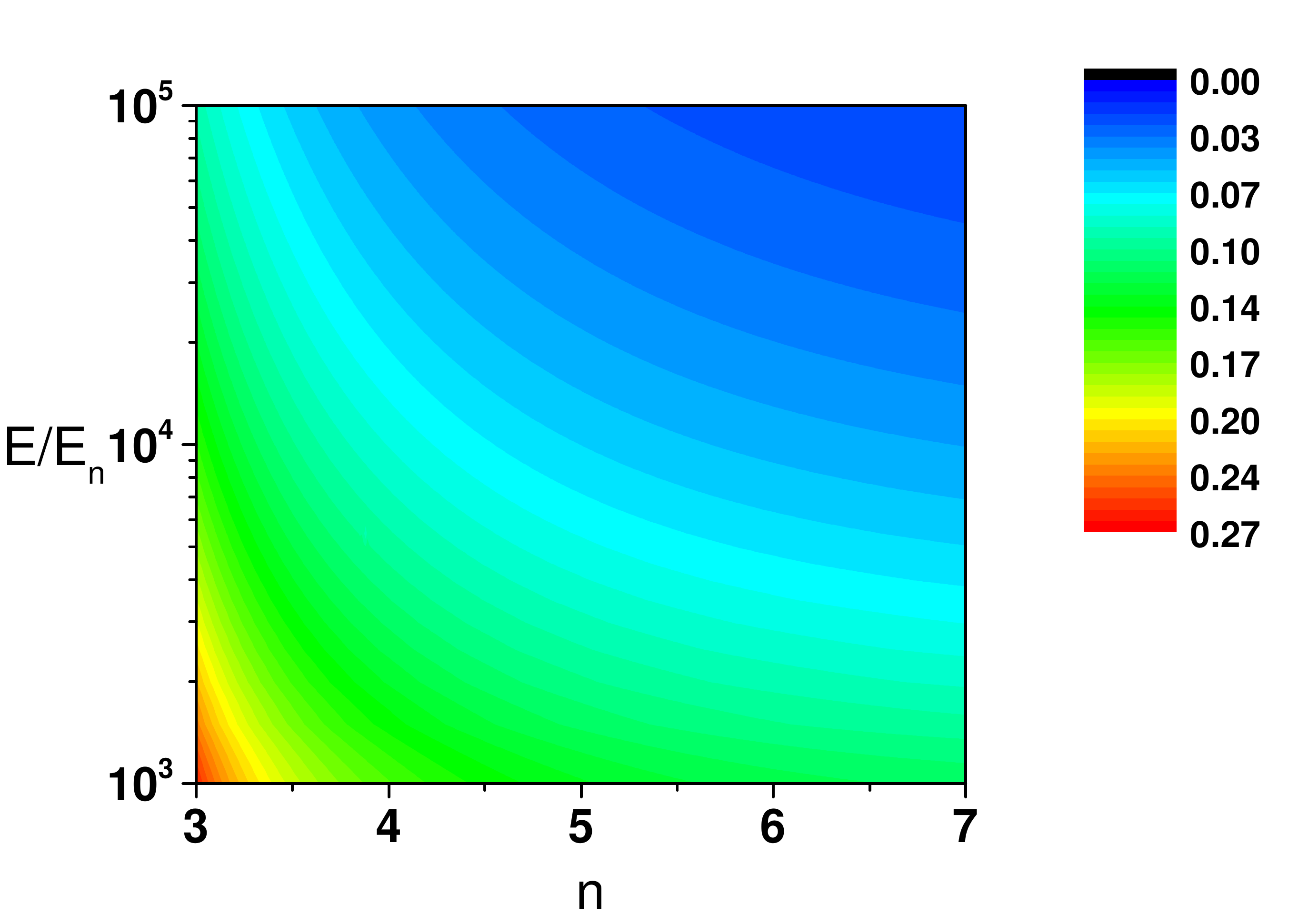}
\caption{
\label{myfigure3}
Relative corrections of reaction rates given by the quantum analytical model with respect to the standard Langevin approach, calculated for different interaction potentials $1/r^n$ and energies for $y=0.01$.}
\end{figure}

The high energy approximations from the previous section do not take into account the presence of shape resonances. If the collision energy is close to the energy of a quasibound state behind the centrifugal barrier, the rates may be significantly modified.  At high temperatures the total effect of many shape resonances in different partial waves should result in some average additional contribution. In this section, we derive a simple model which incorporates this effect. To this end we approximate the centrifugal barrier by an inverted parabolic potential $V(x)=-\frac12 k x^2$, with $k>0$. For such a potential one can find an analytic solution, which is given in terms of parabolic cylinder functions. Considering asymptotic expansions of these solutions at large distances one can show that they have a WKB-like form. We perform the expansion of fully analytical solution far away from the barrier and identify the parts propagating to the left and to the right. After that we calculate the $S$ matrix with the following boundary conditions
\begin{equation}
\begin{split}
\Psi(x) \stackrel{x \to -\infty}{\longrightarrow} &A_{-} \left(\exp\left[ - i \int_{-\infty}^x |k(x')| dx' - i \varphi \right]\right.\\
 &- \left.\frac{1-y}{1+y} \exp \left[ i \int_{-\infty}^x |k(x')| dx' + i \varphi \right] \right)
\end{split}
\end{equation}
at large distances to the left of the barrier, and
\begin{equation}
\begin{split}
\Psi(x) \stackrel{x \to \infty}{\longrightarrow} A_{+} \left( \exp\left[- i \int_x^{\infty} k(x') dx' \right]\right.\\ - \left. S \exp \left[ i \int_x^{\infty} k(x') dx' \right] \right)
\end{split}
\end{equation}
at large distances to the right of the barrier. Here $y$ is as before the parameter describing the reactivity of the system \cite{Idziaszek2010}, $\varphi$ is some arbitrary phase, $S$ denotes the $S$ matrix, $A_{-}$ and $A_{+}$ are normalization coefficients, and $k(x')$ is the local wave vector. Using the exact solution we calculate the $S$ matrix, and the reaction probability $P = 1-|S|^2$. As in this section we are interested only in the behavior of thermally-averaged reaction rates at high temperatures, where the phase shifts vary rapidly with collision energy, we may perform an averaging over the short-range phase. In this way we incorporate the effect of the shape resonances on the reaction rates in an average sense. Calculating reaction probability through the parabolic barrier and performing an average over uniformly distributed values of $\varphi$, we obtain a result valid for arbitrary short-range reaction probability
\begin{equation}
\label{Ppar}
P(\ell,y) = 1-|S_\ell|^2=\frac{P^\mrm{re}}{P^\mrm{re} e^{-2\pi \varepsilon}+1}.
\end{equation}
Here, $\varepsilon=E/\hbar\sqrt{\mu/k}$ denotes dimensionless energy measured with respect to the peak of the parabola. In the universal regime $y=1$ one can recover the WKB solution derived in \cite{Landau}. We can fit analytically the parabolic potential to the centrifugal barrier for arbitrary power-law potential, by equating the first and the second derivative in the maximum of the barrier. Then we change the zero of energy to the asymptotic zero of the physical potential. In the following we will work in dimensionless units defined by $R_n$ and $E_n$.

Applying Eq.~\eqref{Ppar} with the energy
\begin{equation}
\begin{split}
\varepsilon(\ell,E) = &(n/2)^{(2/(n-2))} \frac{1}{\sqrt{2n-4}} E (l(l+1))^{-\frac{n+2}{2n-4}}\\
&- \sqrt{\frac{n}{2}-1}  \frac{\sqrt{l(l+1)}}{n}
\end{split}
\end{equation}
obtained by fitting the parabolic potential to the centrifugal barrier for $V(r)=-1/r^n$, and integrating over angular momenta $\ell$ we get
\begin{equation}
\begin{split}
\label{Kre6}
\mathcal{K}^\mrm{re} \stackrel{E\to\infty}{\longrightarrow}  &g \frac{h}{2 \mu k} \int_0^{\infty} d \ell (2\ell+1) P(\ell,y)\\
 &= g \frac{h}{2 \mu k} \int_0^{\infty} d \ell (2\ell+1) \frac{P^\mrm{re}}{P^\mrm{re} e^{\varepsilon(\ell,E)}+1}
\end{split}
\end{equation}
The reaction probability $P(\ell,y)$ as a function of the continuous variable $\ell(\ell+1)$ is shown in Fig.~\ref{Ferm}. The figure compares the reaction probability calculated from the parabolic potential approximation with the classical approach assuming that only collisions with energies above the barrier contribute to the reaction rate. The latter exhibits a step-like behavior, while the former is reminiscent of a Fermi distribution. The classical description does not include the contribution from the shape resonances, and at the same time overestimates the reaction rate in the regime affected by the quantum reflection. In the universal regime $y=1$ both contributions are almost equal, and in this particular case the Langevin approximation works relatively well. In contrast, for $y<1$ the contribution from the shape resonances is typically larger than the modification due to the quantum reflection above the barrier. In such cases the two effects do not cancel and the Langevin theory underestimates the reaction probability.

Analyzing Fig.~\ref{Ferm} one can develop relatively a simple approximation, allowing one to calculate the integral in Eq.~\eqref{Kre6}. It is based on the observation that the reaction probability $P$ is almost symmetric with respect to the point where $P = \frac12 P^\mrm{re}$, similarly to the Fermi distribution. The integral corresponding to the area below the distribution can be calculated by approximating it by a rectangle
\begin{equation}
K^\mrm{re} \approx g \frac{\pi \hbar}{\mu k} P^\mrm{re} \ell^\ast (\ell^\ast +1),
\end{equation}
where $\ell^\ast$ is the angular momentum corresponding to the point where  $P = \frac12 P^\mrm{re}$
\begin{equation}
\label{P12}
P = \frac{P^\mrm{re}}{P^\mrm{re} e^{ 2 \pi \varepsilon(\ell^\ast,E)} + 1} = \frac {P^\mrm{re}}{2}.
\end{equation}
In the universal case $y=1$ the above equation yields $\varepsilon = 0$, and in this particular case we recover the classical approximation.

In order to verify whether the contribution of quantum corrections due to shape resonances and quantum reflection is still important at high energies, Fig.~\ref{myfigure} depicts the quantum corrections for energies up to $10^6$ $E_n$ for two physically most important power law potentials. We also study the dependence of the quantum corrections on the power $n$ of the potential, for fixed energy and fixed reaction amplitude $y$. They are shown in Fig.~\ref{myfigure2} and~\ref{myfigure3}. Typical energy scales for several systems are shown by Table I.

\begin{table}
\begin{tabular}{|c|c|c|}
\hline
system & power $n$ & $E_n$ [$\mu$K] \\
\hline \hline
$^{174}$Yb$^+$+$^{87}$Rb & 4 & 0.022\\
$^{174}$Yb$^+$+$^{7}$Li & 4 & 3.2\\
He$^{*}$+Ar & 6 & 14000\\
KRb+KRb & 6 & 22.35\\
LiCs+LiCs & 6 & 1.32\\
\hline
\end{tabular}
\caption{Typical energy scales for several systems interacting with atom-ion ($n=4$) or van der Waals ($n=6$) potential at long range.}
\end{table}

\section{Results for van der Waals potential}

\begin{figure}
\centering
\includegraphics[width=\linewidth]{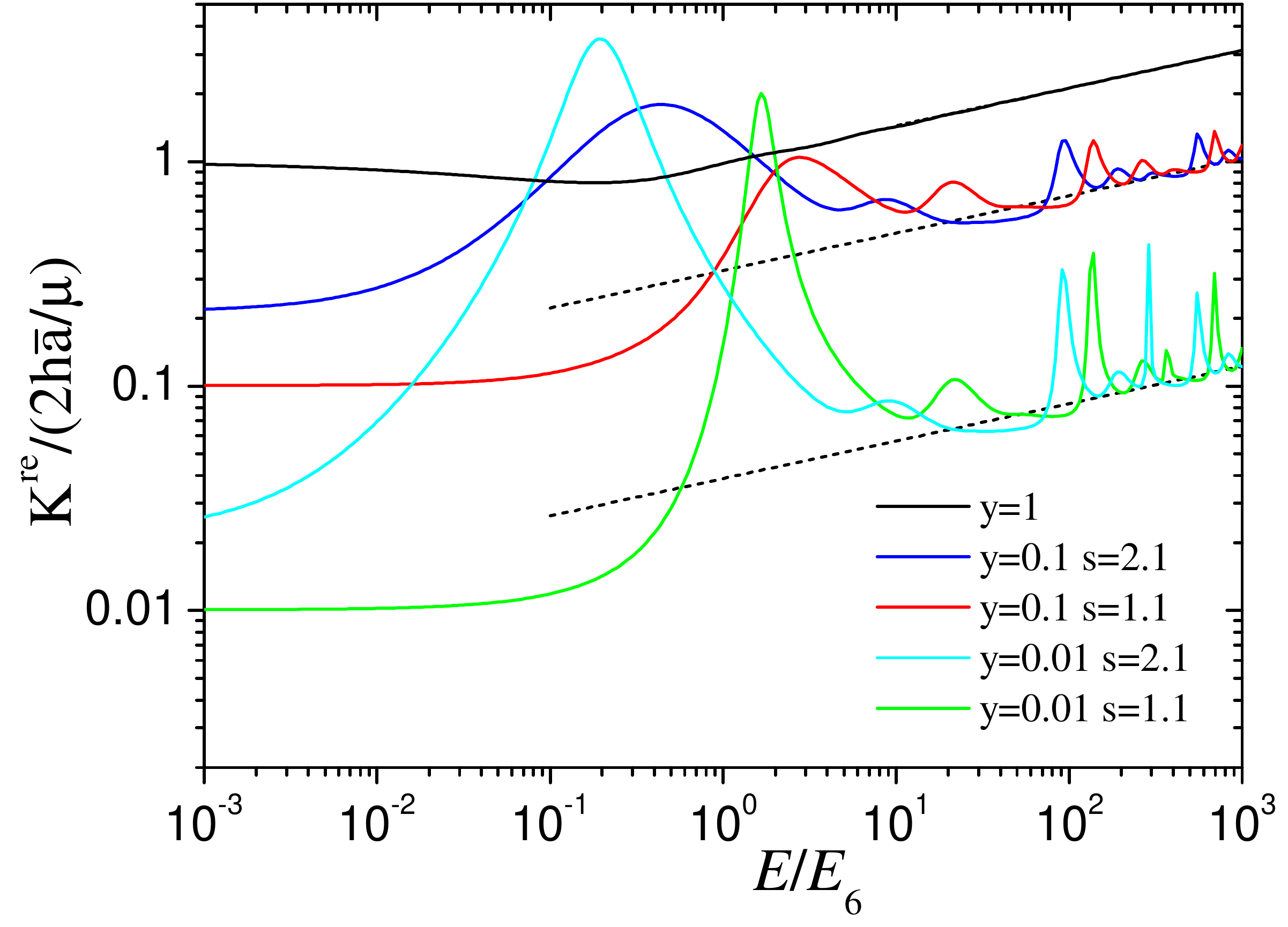}
\caption{\label{KrevsE}(Color online) Reactive rate vs collision energy for distinguishable particles with van der Waals interaction at different reaction amplitudes $y$. The $s$ values are chosen to be close to $p$-wave and $d$-wave shape resonances. The dashed lines show classical approximation~\eqref{KLang}.}
\end{figure}
\begin{figure}
\centering
\includegraphics[width=\linewidth]{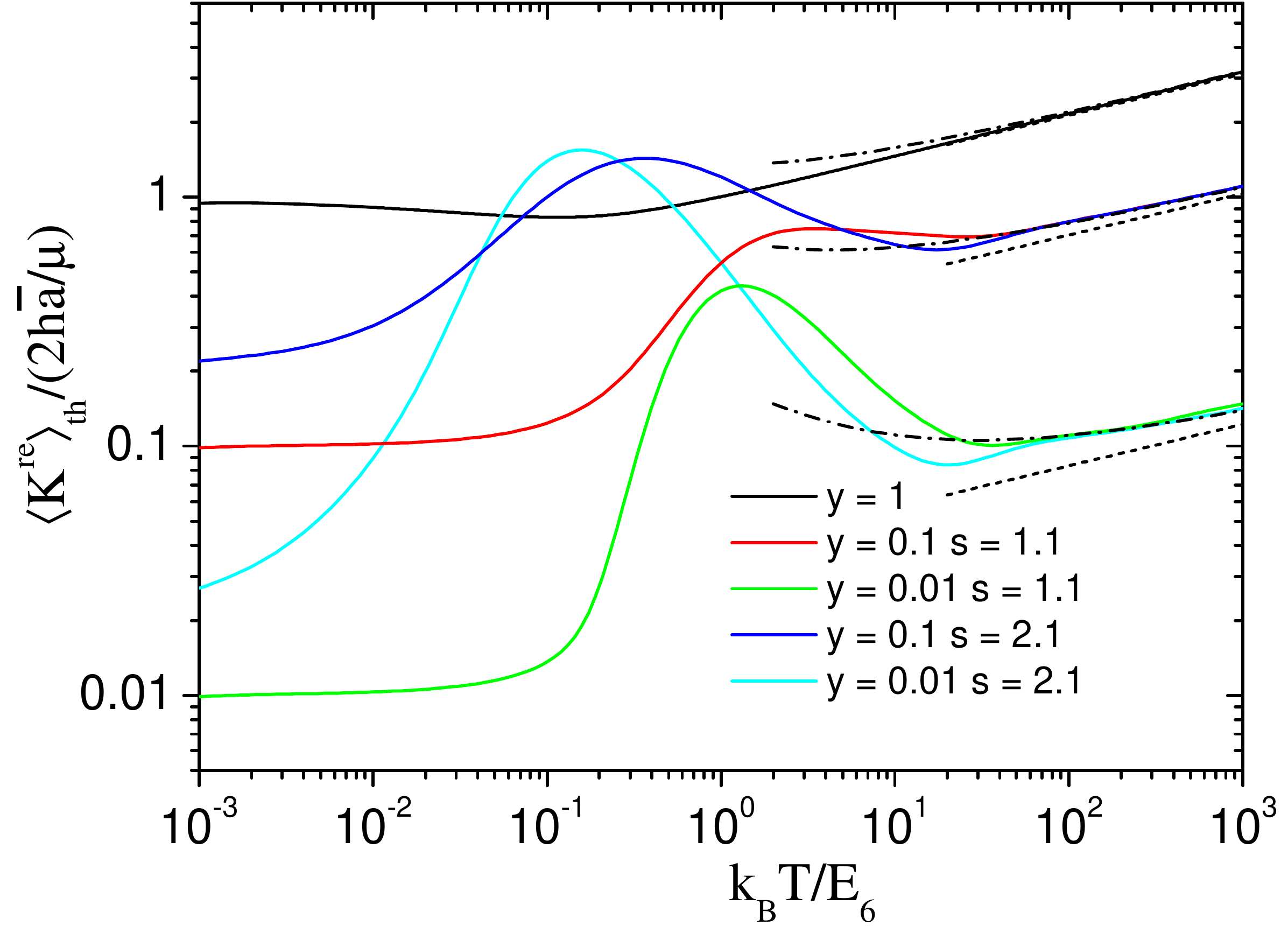}
\caption{\label{KrevsT}(Color online) Same as on Figure~\ref{KrevsE}, but averaged over thermal distribution. The shape resonances can still be seen, especially for low reactivity. The dot-dashed lines show the results obtained using parabolic approximation~\eqref{Kre6}.}
\end{figure}
\begin{figure}
\centering
\includegraphics[width=\linewidth]{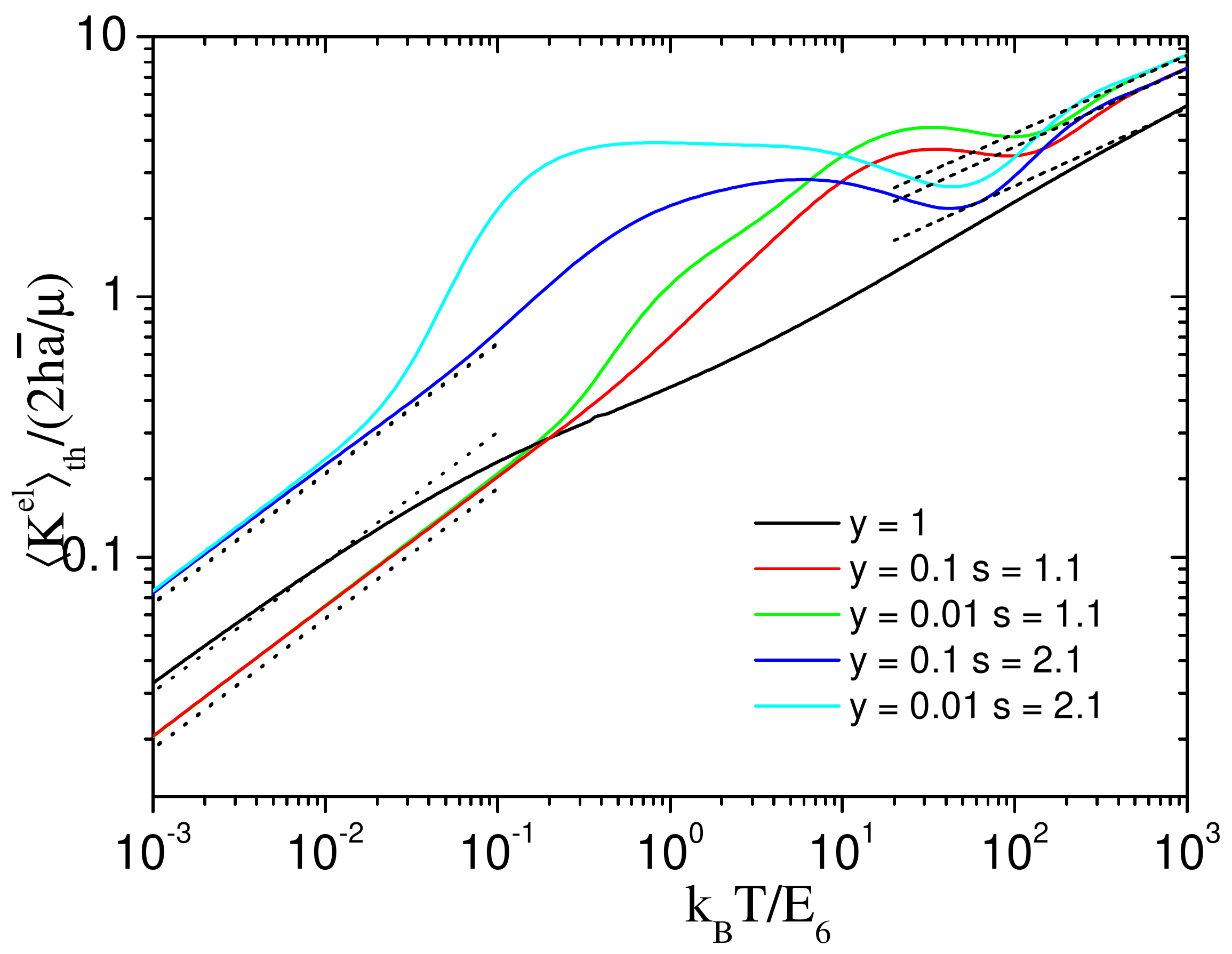}
\caption{\label{KelvsT}(Color online) Thermally averaged elastic rates for the same parameters as on Figures~\ref{KrevsE} and \ref{KrevsT}. The dashed lines show the approximate high energy result given by~\eqref{Kelhigh}. The dotted lines show the low-energy $s$-wave limit. Small discrepancy is due to thermal averaging and $p$-wave contribution.}
\end{figure}

In addition to analytical low- and high energy limits, within our model it is possible to obtain the reactive and elastic rate constants at any collision energy. This can be done either by finding the MQDT functions analytically (for example, in terms of Z functions for van der Waals potential developed by Gao~\cite{Gao1998}) and using formulas~\eqref{KreMQDT} and~\eqref{KelMQDT}, or by numerical treatment. In the latter case we perform scattering calculations, propagating the wave function using Numerov algorithm and extracting the phase shift. The key point here is to set proper boundary conditions at short range, given by~\eqref{Psi}. A possible way to do it is to use solutions of $1/r^n$ potential at zero energy (at short range the kinetic energy is negligible compared to the well depth), combined to reproduce the zero energy limit of the scattering length~\eqref{alm1}. 

At finite energies the reactive rate constant can be greatly enhanced by shape resonances, which are due to the presence of quasibound states behind the barrier. In particular, analytic theory~\cite{Gao1998,Gao2000} predicts a $p$-wave resonance for $s=2$ and a $d$-wave resonance for $s=1$. This is confirmed by the low energy behavior of MQDT functions for those partial waves. The impact of the resonances for near-resonant values of $s$ is presented on Figure~\ref{KrevsE}. We note that the resonances are more important for low values of $y$, where the particles need more time behind the centrifugal barrier for chemical reaction, so forming a quasibound state greatly enhances the reaction rate. At high energies many partial waves contribute to the reaction rate and we observe quite a dense structure of peaks. However, after averaging the reaction rate with respect to thermal distribution $\langle K^{re}\rangle_{th}(T)=2/\sqrt{\pi}(k_B T)^{(3/2)} \int{dE\,\sqrt{E}e^{-E/k_B T}K^{re}(E)}$ the resonances are washed out but on average add an extra contribution to the reaction rate, making it larger than classical approximation~\eqref{KLang}, as can be seen on Figure~\ref{KrevsT}. At energies above $\sim 100E_6$ the parabolic approximation starts to agree well with the numerical results, giving a good estimate of this contribution.

The elastic rate is particularly important for experiments which aim to use the evaporative cooling technique~\cite{Stuhl2012}. Reaching thermal equilibrium is possible only if the elastic collisions are more frequent than chemical reactions. Formulas~\eqref{KLang} and \eqref{Kelhigh} predict that at high energies the elastic rate behaves like $E^{3/10}$, while the reactive one like $E^{1/6}$, so elastic collisions should dominate for hot gases, but not necessarily in the evaporative cooling regime. Figure~\ref{KelvsT} shows the elastic rate for some exemplary cases. The high energy approximation~\eqref{Kelhigh} agrees with exact calculations at energies above $\sim 100E_6$.

\section{Conclusions}
We introduced a simple model of a reactive collision basing on the formalism of quantum defect theory. We represented the inelastic processes by a single, strongly open collision channel. Our model can be applied to all systems for which the long range interaction behaves like $1/r^n$ and describes the collision by two parameters: $y$, connected with short-range probability of reaction, and $s$, describing the phase shift. We obtained analytical formulas for the low energy limits of elastic and reactive rates in terms of those parameters. We also discussed the behavior of the rates at finite temperatures and derived their high energy limits. Our theory takes into account the effect of shape resonances, which may increase the reaction rate above the universal values and explains the observed scattering resonances in collisions of argon with metastable helium~\cite{Jachymski2013}.

For realistic systems one can expect more terms in the interaction potential, such as $C_8/r^8$ and higher order ones, small exchange terms $\propto r^{-3}$ and others. In many cases they have negligible contribution at distances $\sim R_n$ and thus can be incorporated in the short range boundary conditions, affecting only the $s$ parameter. They can, however, influence the dynamics especially at energies much smaller or much larger than $E_n$. In this case, numerical treatment using MQDT boundary conditions is still possible (see e.g.~\cite{Jankunas2014}). We expect that using only single van der Waals term in the potential and constant $y,s$ parameters should work better for heavier systems. In general, the $y,s$ parameters can fluctuate with energy and partial wave; especially $s$, which is connected with the phase shift, can change. Another possible numerical approach, also taking advantage of the short range nature of interchannel couplings and using MQDT to obtain scatteirng properties from short range K matrix, is presented in~\cite{Hazra2014}.

We also did not consider here the effect of multiple closed channels, which introduce additional resonance effects. In fact the density of closed channel states may be very high and in some physical systems one should expect multiple overlapping resonances~\cite{Mayle2012}, which our simple model cannot reproduce. In this case the particles form a collision complex with large phase space and effectively ``stick'' to each other for long times. Interestingly, in the highly resonant regime it is reasonable to make statistical assumptions about the strength of the interchannel couplings basing on Gaussian Orthogonal Ensemble. Within this model the collision complex, once created, ergodically explores the available phase space. In the limit of many possible exit channels, this brings the reaction rate back to the universal limit~\cite{Mayle2013}, as it is impossible to come back to the entrance channel and thus there is no outgoing flux from the short range. It is also possible that the underlying physics is in fact controlled by a few dominating resonances, while most of the other ones have negligible impact on the collision process (for example, in the collision of two cesium atoms most of the high partial wave resonances are extremely narrow~\cite{Berninger2013}, so they would not contribute much to the collision rates). Exploring this situation will be the subject of our future research.

\section{Acknowledgments}
We thank Manuel Lara for interesting discussions. This work was supported by the Foundation for Polish Science International PhD Projects and TEAM programmes co-financed by the EU European Regional Development Fund, National Center for Science grants No. DEC-2011/01/B/ST2/02030, DEC/2012/07/N/ST2/02879 and DEC-2013/09/N/ST2/02188.
\bibliography{Allrefs}
\end{document}